\documentstyle[multicol,aps,prl,epsf]{revtex}
\begin{document}

\draft

\title{
Flat-band ferromagnetism proposed for an organic polymer crystal
}

\author{
Ryotaro Arita$^1$, Yuji Suwa$^2$, Kazuhiko Kuroki$^3$, 
and Hideo Aoki$^1$
}

\address{$^1$Department of Physics, University of Tokyo, Hongo,
Tokyo 113-0033, Japan}
\address{$^2$Advanced Research Laboratory, Hitachi Ltd., 
Hatoyama, Saitama 350-0395, Japan}
\address{$^3$Department of Applied Physics and Chemistry,
University of Electro-Communications, Chofu, Tokyo 182-8585, Japan}

\date{\today}

\maketitle

\begin{abstract}
Motivated from the flat-band ferromagnetism conceived theoretically 
for a single chain of five membered rings (polyaminotriazole) 
by Arita {\it et al.}, [Phys. Rev. Lett. 
{\bf 88}, 127202 (2002)], we have studied whether the magnetism 
can indeed occur as a bulk, i.e., in a three-dimensional crystal 
of the polymer, by means of the spin density functional calculation. 
We find that the intra-chain 
ferromagnetism is robust against crystallization as far as 
the flat band is made half-filled.  
We have further investigated the actual crystal doped with 
various compounds, where HF$_2$ is shown to put the system 
close to the bulk ferromagnetism while stronger anions 
such as BF$_4$ or PF$_6$ should be promising.
\end{abstract}

\medskip


\begin{multicols}{2}
\narrowtext
Materials design, especially that from the standpoint of  
many-body effect, is one of the most challenging avenues 
in condensed matter physics. In particular, designing 
magnets in materials consisting entirely of non-magnetic 
elements is an issue of great interest. In fact, a variety 
of ideas for exotic ferromagnetism have been proposed 
in this decade, where first-principles calculations are combined 
with electron-correlation studies for models exemplified by the 
Hubbard model\cite{Shima93,Okada00,Arita98,Arita02,Kusakabe02}. 
In this context, organic ferromagnets are of special interest.  
While organic ferromagnetism has been realized in 
localized-spin systems such as TDAE-C$_{60}$\cite{Allemand91}, 
whether we can have a {\it band (i.e., itinerant) 
ferromagnetism} in organics remains a challenging question.  

Recently, we have proposed a new possibility of itinerant 
ferromagnetism in an organic {\it polymer} of five-membered 
rings, polyaminotriazole (PAT), and have explained the origin 
of the magnetism in terms of Mielke and Tasaki's flat-band 
ferromagnetism\cite{Tasaki-review} by coupling a first-principles 
band calculation with a study for the Hubbard 
model\cite{Arita02}. Concisely, the {\it flat-band ferromagnetism} 
occurs when there is a dispersionless band which satisfies a special 
condition (called connectivity condition) in the one-electron band 
structure. Note that the magnetism is distinct from the usual 
narrow-band limit in that the magnetism occurs as a quantum 
interference effect in the repulsively interacting system 
for a {\it finite} transfer energy. 

While the monomer (aminotriazole) is commercially available, 
experimental attempts at its polymerization have just
begun\cite{Nishihara02}.  Experimentally, we have to aim at 
a bulk magnetization, so theoretically it is imperative 
to study whether we can have a ferromagnetism in three-dimensional 
crystals, which is of fundamental as well as practical interests.  

The key issues are
(i) the robustness of the intra-chain ferromagnetism to start with 
(i.e., whether the inter-chain interaction in the crystal 
disturbs the intra-chain ferromagnetism), 
(ii) whether the inter-chain interaction can be ferromagnetic (i.e., 
the competition between the
ferromagnetic (direct-exchange) inter-chain interaction and the 
inter-chain energy gained by the antiferromagnetic order), 
and their energy scale, 
and (iii) which chemial dopants should be opted to 
bring the flat-band close to half-filling, a necessary 
condition for the flat-band ferromagnetism.  
The present paper addresses these problems.  

Problem (i) is non-trivial, since the presence of adjacent chains 
can affect the magnetism within the chain\cite{Ruini02}. 
Conversely, if the inter-chain 
interaction does not destroy the intra-chain ferromagnetism, 
this is already interesting for experimental opportunities, since 
if we can for example epitaxially grow the organic crystal on a 
metallic substrate with a 
large work function such as platinum, the 
interface state of the PAT chain can possibly have a net polarization.
In fact, we shall show that the presence of neighboring chains 
in the crystal does not destroy the intra-chain ferromagnetism 
for the realistic chain-chain distance. 
 
On the other hand, problem (ii) on the inter-chain magnetic coupling, 
we shall show that ferromagnetic and antiferromagnetic 
interactions are both small in magnitude and their competition 
is subtle. 

As for (iii), we we have studied various anions as dopants in PAT crystal 
to examine the magnetic property.  
We shall conclude that, while the dopant studied here, ClO$_4$, F, 
and HF$_2$, bring the system close to, but not in, the 
ferromagnetic phase, the chemical 
trend indicates that stronger anions having higher electron 
affinity (BF$_4$ and PF$_6$) are 
expected to realize the flat-band ferromagnetism. 

We have adopted a first-principles band calculation within the 
framework of the generalized gradient approximation based on the 
spin-density functional theory (GGA-SDFT) to 
compare the total energy of various magnetic states.  
In the GGA-SDFT calculations 
the exchange-correlation functional introduced by 
Perdew, Burke, and Wang\cite{Perdew1996} is adopted with the ultra-soft 
pseudopotential\cite{Vanderbilt90,Laasonen93} in a separable form. 
The wave functions are expanded by plane waves up to a cut-off energy 
of 20.25 Ry.

To clarify the physics the doping has been done in two ways. 
We first look at the band-filling dependence by reducing the number 
of electrons, where a uniform negative background charge is introduced to 
make the system charge-neutral as in Ref.\cite{Arita02}. 
The latter part of the paper examines actual anions as dopants, 
where the charge transfer across the anion and PAT is calculated 
self-consistently. 

\begin{figure}
\begin{center}
\leavevmode\epsfysize=60mm \epsfbox{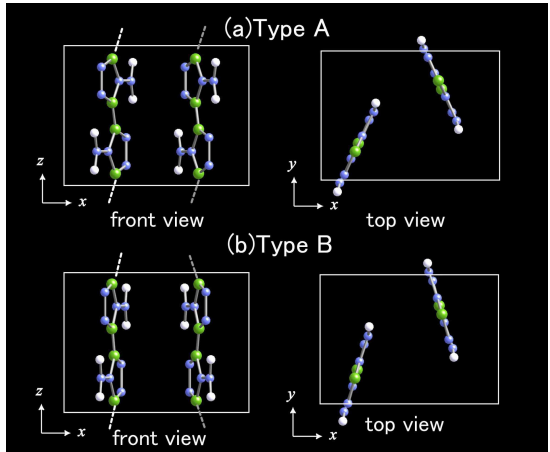}
\caption{The optimized
atomic configuration of crystallized polyaminotriazole 
(before doping) for type A and type B.
The green, white, blue balls represent C, H and N, respectively, 
and the square represents the unit cell.
}
\label{Xtaluniform}
\end{center}
\end{figure}

We start with the atomic configuration of the 
(undoped) PAT crystal obtained by the 
structure-optimization, shown in Fig.\ref{Xtaluniform}, where 
the total energy is minimized with the conjugate gradient 
scheme\cite{Yamauchi1996}. We have assumed that the unit-cell 
is orthorhombic and contains two chains of PAT which form a herringbone 
structure. We have studied two types of the crystal structure, 
A(non-staggered) and B(staggered).  Similar crystal structures 
have been considered in recent calculations for 
polythiophene\cite{Bussi02} and poly-phenylene-vinylene\cite{Ruini02}, 
where PT2 and PT1 in Ref.\cite{Bussi02} correspond to our A and B, 
respectively. 

We have determined the size of the unit-cell as follows.
We first determined the dimension of the unit-cell along 
the chain to be 7.17 \AA, which minimizes the ground state energy 
of an isolated chain\cite{Arita02}. 
For the unit-cell size perpendicular to the chain, we have 
calculated the total energies by changing the linear dimension 
with the increment of 1.0 au (i.e., $0.529$ \AA) 
to search for the energy minimum.
The resulting unit-cell size is 8.46\AA $\times$ 6.35\AA.

We have then doped the system, first by reducing the number of electrons.  
We focus on the case of the 
half-filled flat band, for which the flat-band ferromagnetism is 
originally conceived.  In the present crystal, 
top four valence bands are nearly flat to be precise (corresponding 
to the presence of four five-membered rings in a unit cell), 
and the half-filling refers to the case of one carrier per ring.  
We have obtained four types of solution for 
both A and B crystal structures in the doped case: 
an intra-chain ferromagnetic(F) and inter-chain antiferromagnetic(AF) 
state (which we call interAF hereafter), 
an intra-chain AF and inter-chain F state 
(intraAF), a state antiferromagnetic both for intra-chain and inter-chain 
(AF$^2$), and finally the ferromagnetic state.   

\begin{figure}
\begin{center}
\leavevmode\epsfysize=80mm \epsfbox{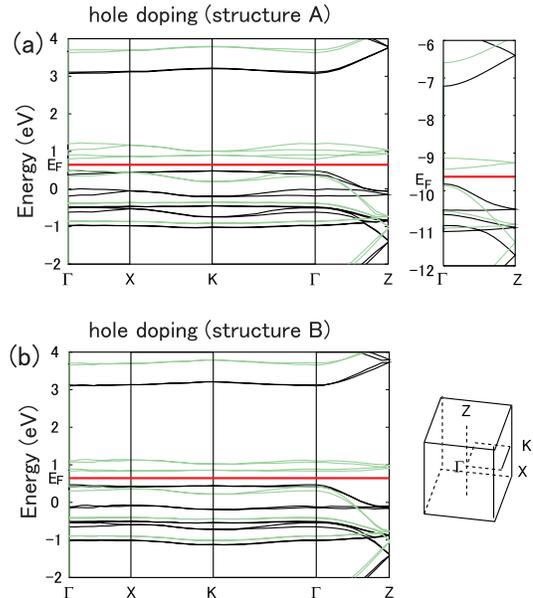}
\caption{The band structure for the ferromagnetic 
solution for the hole-doped 
crystallized polyaminotriazole in type A(a) and 
type B(b) structure. 
The black (green) lines represent the bands for the
majority (minority) spin.
Top-right inset shows the band structure for 
an isolated chain of doped polyaminotriazole.
}
\label{Banduniform}
\end{center}
\end{figure}

The band structure for the doped system is shown in Fig.\ref{Banduniform} 
for the ferromagnetic solution. 
For both A and B structures, we can see that the 
dispersion along the chain is quite similar to that of an isolated 
PAT chain (shown in the inset), from which we can see that the 
effect of the inter-chain interaction is small for the ferromagnetic state.

The total energies of the three AF states for the structure A(B), 
as measured from those of the F state in respective structures, are, 
respectively : 
$E({\rm interAF})=-40$ (0) meV, $E({\rm intraAF})=130$ (100) 
meV and $E({\rm AF^2})=40$ (100) meV. 
If we estimate the ``intra-chain magnetic coupling'' 
as $J_{\rm intra} = ({\rm intraAF})-E({\rm F})$, 
and ``inter-chain magnetic coupling'' as 
$J_{\rm inter} = E({\rm interAF})-E({\rm F})$, 
$J_{\rm intra}$ is $\sim -100$ meV ($<0$, i.e., ferromagnetic) for 
both A and B structures. 
If we compare this value with that for a single PAT chain 
(estimated to be $\sim -50 \times 2=-100$ meV\cite{Arita02}, 
where the factor 2 is required for comparison with the present value 
for two chains per unit cell), 
the intra-chain ferromagnetism is seen to be 
quite robust even in the crystal.

On the other hand, the situation is subtle for $J_{\rm inter}$.  
While $J_{\rm inter}$ is estimated to be 40 meV 
($>0$, i.e., AF) for 
structure A, $J_{\rm inter}\sim 0$ meV (paramagnetic) for B. 
This implies that F and AF 
interactions are both small in magnitude, so that the value of 
$J_{\rm inter}$ should depend on details of the crystal structure.

\begin{figure}
\begin{center}
\leavevmode\epsfysize=65mm \epsfbox{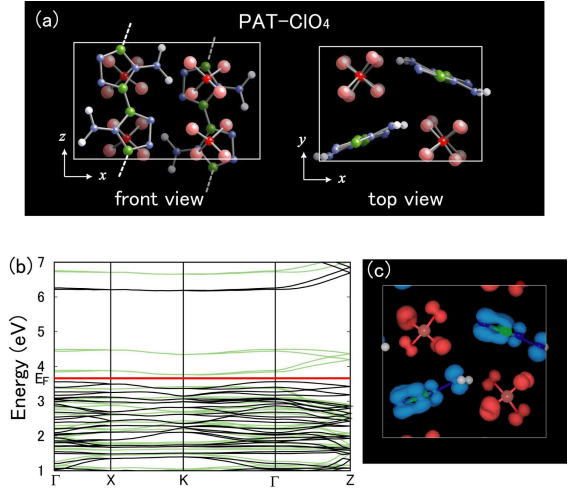}
\caption{
(a) The optimized atomic configuration of the crystallized
polyaminotriazole doped with ClO$_4$.  
The green, white, blue, red, pink, orange balls represent the 
position of the C, H, N, O, Cl and F atoms, respectively.  
(b) The band structure for the ferromagnetic solution. 
The black (green) lines represent the bands for the
majority (minority) spin.
(c) The wave function (sum of the squared absolute values of 
the top-four, majority-spin valence bands 
at $\Gamma$) for the ferromagnetic polyaminotriazole crystal 
doped with ClO$_4$.  Blue (red) contours represent 
the amplitude around PAT (dopant). 
}
\label{ClO4}
\end{center}
\end{figure}

This is why we move on to first-principles calculation of 
the flat-band ferromagnetism for actual chemical dopants introcuded 
in PAT crystal.  
We should opt for anions having large electron affinities, 
because the bands of the dopants to which the charge transfer 
from PAT occurs will then lie well away from the bands around the 
Fermi energy (i.e., the flat bands).  
In the present study, we have studied halogens (Cl, F) and halogen 
compounds (ClO$_4$, HF$_2$). To make the flat bands half-filled,
we put four dopants 
in each unit-cell that contains four five-membered rings.  
As for the polymorph we focus here on structure A, and  
we have adopted a slightly larger unit-cell size  
(9.52\AA $\times$ 7.41\AA $\times$ 7.17\AA) 
to accommodate ClO$_4$ and HF$_2$.

We first found that we can exclude the Cl-doped case 
since the paramagnetic solution has a total energy lower 
than that for the ferromagnetic one by $\simeq$ 270 meV.  
For ClO$_4$, F and HF$_2$, on the other hand, 
the ferromagnetic state becomes lower in energy than 
the paramagnetic state. In Fig.\ref{ClO4}(a), we show the 
optimized atomic configuration for the 
ferromagnetic solution in the ClO$_4$-doped crystal.  
We can see that the insertion of the ClO$_4$ molecules 
significantly changes the relative position of the two PAT chains 
in the unit-cell from those in Fig.\ref{Xtaluniform}, 
which suggests that the molecules may
crucially affect the electronic properties and unfavors 
the intra-chain ferromagnetism.

So let us turn to the band structure in Fig.\ref{ClO4}(b) 
to see whether the states around $E_F$ are modified.
If the electron affinity of ClO$_4$ were sufficiently larger than that 
of PAT, the ClO$_4$-charactered bands would not hybridize with the 
states around $E_F$. However, the result shows that 
the hybridization occurs for the flat (valence-top) bands, 
and the Mielke-Tasaki's orbits are severely affected. 

\begin{figure}
\begin{center}
\leavevmode\epsfysize=65mm \epsfbox{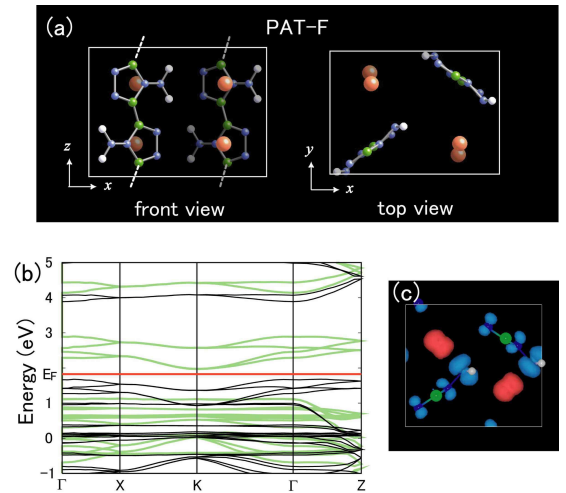}
\caption{A plot similar to Fig.3 for the fluorine-doped case.
}
\label{F}
\end{center}
\end{figure}

For a more quantitative argument, we have looked at 
the charge distribution by adopting a 
method due to Aizawa and Tsuneyuki\cite{Aizawa,Suwa02b} 
for estimating the Mulliken charge 
for plane-wave basis band calculations. 
We divide the whole system into subsystems with each subsystem 
containing one atom, and calculate the valence charge for each.  
Summing the valence charges for dopant (ClO$_4$ here), 
we can estimate the charge transfer from PAT to the dopant($Q$).  
For PAT-ClO$_4$, we have found that $Q$ 
for each dopant molecule (i.e., for each five-membered ring) 
is $\simeq 0.88$, i.e., the charge transfer is incomplete.  
This suggests that the bands 
(Mielke-Tasaki's) to which we wanted to hole-dope 
are hybridized with the dopant (ClO$_4$) band.  
We can indeed confirm this from the plot of the 
(majority-spin) wave function 
as a sum ($\rho$ hereafter) of the squared absolute values of 
flat bands at $\Gamma$ in Fig.\ref{ClO4}(c). 
We can see that the amplitude extends substantially 
to ClO$_4$, especially on O sites.
In fact, the ground state becomes antiferromagnetic even within the chain, 
with  $J_{\rm intra}=60$ meV and $J_{\rm inter}=20$ meV.

Let us now move on to the fluorine-doped case. 
We show the optimized atomic 
configuration and the band structure in Fig.\ref{F}. 
The band structure is 
similar to those plotted in Fig.\ref{Banduniform} with 
the charge transfer estimated to be $Q \simeq 0.78$. 
The wave function in Fig.\ref{F}(c) indicates 
that the amplitudes on C atoms becomes considerably small, 
which means that the connectivity 
condition, necessary for the flat-band ferromagnetism\cite{Tasaki-review}, 
is violated.
The exchange energies are estimated to be 
$J_{\rm intra}= 40$ meV and 
$J_{\rm inter} = 20$ meV, so that the ground 
state is antiferromagnetic both within and 
across the chains.

\begin{figure}
\begin{center}
\leavevmode\epsfysize=65mm \epsfbox{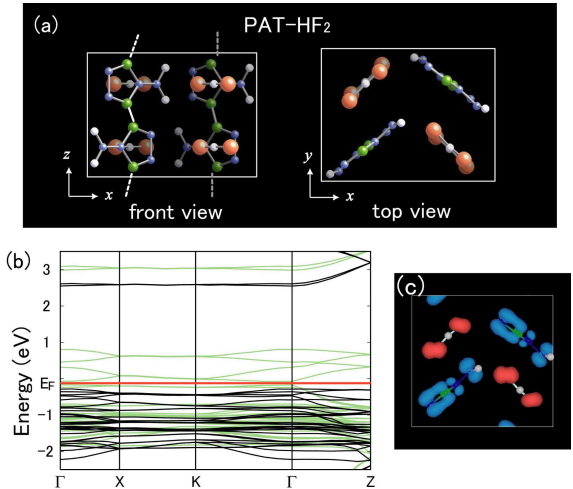}
\caption{A plot similar to Fig.3 for the HF$_2$-doped case.
}
\label{HF2}
\end{center}
\end{figure}

Let us finally discuss the result for HF$_2$, which has the 
largest electron affinity among the dopants studied here. 
While the band structure and wave function in 
Fig.\ref{HF2} suggest that 
Mielke-Tasaki's states hybridize with anion states to 
some extent, the charge transfer is calculated to be $Q=1.1$, 
so that this case is more promising than that of ClO$_4$ or F.
While $J_{\rm intra}\simeq 10$ meV and
$J_{\rm intra}\simeq 30$ meV are still both 
antiferromagnetic, $J_{\rm intra}$ becomes 
considerably smaller than those for 
ClO$_4$ and F, reflecting the high electron affinity of HF$_2$.  
From the chemical trend, we may expect that 
a favorable situation (similar to the uniform doping described in 
the first half of the paper) for the ferromagnetism 
is expected to be realized for BF$_4$ or PF$_6$, which are known 
to have higher electron affinity, although 
a first-principles calculation for BF$_4$ or PF$_6$ would be 
too demanding, since these have many valence 
electrons or require large cut-off energies 
in the plane-wave expansion. 

To summarize, a spin density functional calculation shows that the
three-dimensional crystal of chains of five-membered rings 
has a robust intra-chain ferromagnetism originally conceived 
for an isolated polymer.  
For the chemical doping 
BF$_4$ or PF$_6$ with high electron affinity should be promising.  
The magnetism in the organic crystal 
may also be controlled by the pressure effect.  

We would like to thank H. Nishihara, Y. Yamanoi and S. Nakao for fruitful 
discussions. This work was supported in part by a Grant-in-Aid 
for Creative Research (No.14Gs0297)
and Special Coordination Funds for Promoting Science and
Technology from Japanese Ministry of Education. 
The GGA calculation was performed with 
TAPP (Tokyo Ab-inito Program Package), where RA and YS would like 
to thank J. Yamauchi for useful discussions. Numerical calculations 
were performed on SR8000 in ISSP, University of Tokyo.

\end{multicols}
\end{document}